\documentclass[floatfix,preprint,aps,prd,showpacs,amsmath,amssymb]{revtex4}
\usepackage{graphicx}
\usepackage{bm}

\newcommand{\mr}[1]{\mathrm{#1}}

\newcommand{\cgw}[0]{c_{\mr{gw}}}
\newcommand{\cem}[0]{c_{\mr{em}}}
\newcommand{\rau}[0]{R_{\mr{au}}}

\begin{document}

\title{A R{\o}mer time-delay determination of the 
gravitational-wave propagation speed}

\author{Lee Samuel Finn}
\email{LSFinn@psu.edu}
\affiliation{Department of Physics,
The Pennsylvania State University, State College, PA, 
USA 16802-6300}
\author{Joseph D.\ Romano}
\email{joe@phys.utb.edu}
\affiliation{Department of Physics and Astronomy, 
University of Texas at Brownsville, 
Brownsville, TX 78520}

\date{1 April 2013}


\begin{abstract}
In 1676 Olaus R{\o}mer presented the first observational evidence for a finite light velocity $\cem$. He formed his estimate by attributing the periodically varying discrepancy between the observed and expected occultation times of the Galilean satellite Io by its planetary host Jupiter to the time it takes light to cross Earth's orbital diameter. Given a stable celestial clock that can be observed in gravitational waves the same principle can be used to measure the propagation speed $\cgw$ of gravitational radiation. Space-based ``LISA''-like detectors will, and terrestrial LIGO-like detectors may, observe such clocks and thus be capable of directly measuring the propagation velocity of gravitational waves. In the case of space-based detectors the clocks will be galactic close white dwarf binary systems; in the case of terrestrial detectors, the most likely candidate clock is the periodic gravitational radiation from a rapidly rotating non-axisymmetric neutron star. 
Here we evaluate the accuracy that may be expected of such a R{\o}mer-type measurement of $\cgw$ by foreseeable future space-based and terrestrial detectors. 
For space-based, LISA-like detectors, periodic sources are plentiful: by the end of the first year of scientific operations a LISA-like detector will have measured $\cgw$ to better than a part in a thousand. Periodic sources may not be accessible in terrestrial detectors available to us in the foreseeable future; however, if such a source is detected then with a year of observations we could measure $\cgw$ to better than a part in a million. 
\end{abstract}


\maketitle


\section{Introduction}\label{sec:intro}

%
%

Over the course of a Jovian synodic year the distance light must transit 
between Earth and Jupiter varies by two astronomical units: 
approximately $3\times10^8$~km. 
If we neglect the time required for light to cross Earth's orbit, 
the interval between events that are periodic at Jupiter will at 
Earth be observed at times that may vary from the expected by 
as much as $10^3$~s. 
It was this observed variation between the observed and expected 
occultations of the Galilean satellite Io that led Olaus R{\o}mer 
to conclude that light has a finite propagation speed and to the
first real measurement of light's 
velocity $\cem$ \cite{rmer:1676:dtm,rmer:1677:dcm}. 
Galactic compact white dwarf binary systems, or rapidly rotating 
non-axisymmetric neutron stars, are similarly regular clocks whose 
periodic signal is transmitted to Earth via gravitational radiation. 
In the same way that R{\o}mer was able to use observations of
the discrepancies in the optically measured times of Io's 
occultations by Jupiter to measure the speed of light, 
so we may use 
gravitational-wave observations by space-based LISA-like detectors \cite{danzmann:2011:luh,baker:2011:shl,baker:2011:sml,baker:2011:sll,baker:2011:sll:1}
of compact white dwarf binary systems, or by terrestrial LIGO-like 
detectors \cite{harry:2010:aln,accadia:2011:sov,somiya:2012:dco,hild:2011:ssf} of rapidly rotating neutron stars, to measure the propagation 
speed $\cgw$ of gravitational waves. 

In general relativity theory gravitational waves and light wave both propagate on spacetime null geodesics; correspondingly, there is no difference in their respective (vacuum) propagation speeds. A direct measurement of the gravitational-wave propagation speed is, thus,  a ``go/no-go'' test of the theory. 

Measurement of the R{\o}mer delay directly and unambiguously access the wave propagation speed across Earth's orbital baseline. This stands in contrast to other proposed tests of general relativity whose results are sometimes discussed in terms of the gravitational-wave propagation speed, but whose interpretation in this way requires a theoretical model or phenomenological framework to relate the observation to the propagation speed. For example, \citet{will:1998:bmo} suggested searching for an anomalous (compared to general relativity's prediction) compression of the gravitational-wave signal from an inspiraling binary system. Such a compression could be interpreted as a frequency-dependent gravitational-wave propagation speed resulting from a non-zero ``graviton-mass''. In a similar vein, \citet{larson:2000:ubs,cutler:2003:lbs} proposed measuring the phase difference between the binary's orbital phase at some fiducial time as determined optically and by gravitational-wave observations. The phase difference, relative to that predicted by general relativity, could then be interpreted as differences in the gravitational-wave propagation speed at a frequency twice the binary orbital frequency. What is relevant is that, unlike the measurement described in this work, none of these other measurements directly accesses the gravitational-wave propagation speed: i.e., their interpretation in terms of the wave propagation speed requires a theory or phenomenological framework that relates the observed phenomena to the wave propagation speed. As shown by \citet{carlip:2004:mos} in the context of a recent claim to have measured the propagation speed of the gravitational force \cite{fomalont:2003:mol} through its effect on light travel, change the theory or framework and the interpretation changes. 

In Section \ref{sec:measurement}
we estimate the precision to which we can measure 
the gravitational-wave propagation speed 
from multi-year observations of periodic 
gravitational waves.
We assume here that the gravitational-wave 
frequency and sky location of the source
are known \emph{a priori}.
For such cases, we show that 
the Fisher matrix estimate of 
the uncertainty can be expressed very simply 
in terms of the source parameters and orbital 
radius of the Earth's motion around the Sun, 
valid for \emph{all} ground-based 
and proposed space-based detectors.
Details specific to the detectors, such as 
antenna pattern functions, cancel out when the 
uncertainty is expressed in terms of the signal-to-noise 
ratio of the measurement.
We also discuss the complications introduced if
we relax the assumption of \emph{a~priori} knowledge
of the source frequency and sky location of the 
gravitational-wave source.
Although the calculation is more complicated for 
this case, the Fisher matrix formalism can still be
used to estimate the uncertainty in the gravitational-wave
propagation speed as a function of the source parameters 
and detector geometry. 
In Section \ref{sec:discussion} we use 
the general result of Section~\ref{sec:measurement}
to obtain numerical values 
for ``$3\sigma$" fractional uncertainties in $\cgw$
for observations in terrestrial and space-based detectors.
We also summarize our conclusions.



\section{Estimating the measurement precision of the 
gravitational-wave propagation speed}
\label{sec:measurement}

\subsection{Introduction}

We use the Fisher Information Matrix formalism 
\cite{finn:1992:dma,vallisneri:2008:uaa} to estimate 
the precision with which we can estimate the 
gravitational-wave propagation speed from 
multi-year observations 
of periodic gravitational waves whose 
frequency and propagation direction are known \emph{a priori}. 
For example, in the case of observations in a 
ground-based detector or detector network, the source may 
be a rapidly rotating neutron star whose rotational frequency and sky location are known from observation of its radio pulses; or, in the case of observations 
made with a space-based detector, the source 
may be a close white dwarf binary system that has also been observed optically. 

For any monocrhomatic source we may write the TT gauge gravitational wave strain at time $t$ and location $\vec{x}$ as
\begin{subequations}
\begin{align}
\bm{h}(t,\vec{x}) &= h_{+}(t,\vec{x})\bm{e}_{+} + h_{\times}(t,\vec{x})\bm{e}_{\times}\,,
\end{align}
where $\bm{e}_{+}$ and $\bm{e}_{\times}$ are orthogonal polarization tensors, fixed in inertial space, and
\begin{align}
h_{+}(t,\vec{x}) &= H_{+} 
\cos\left[2\pi f_{\text{gw}}u+\Phi_{+}\right]\,,\\
h_{\times}(t,\vec{x}) &= H_{\times}
\cos\left[2\pi f_{\text{gw}}u+\Phi_{\times}\right]\,,\\
u &= t-\frac{\hat{k}\cdot\vec{x}}{\cem(1+\epsilon)}\,.
\end{align}
\end{subequations}
Here $\epsilon$ is the fractional difference between light 
speed ($\cem$) and the gravitational-wave propagation 
speed ($\cgw=\cem(1+\epsilon)$), $\hat{k}$ is the 
wave-propagation direction and 
$H_{+}$, $H_{\times}$, $\Phi_{+}$ and $\Phi_{\times}$ 
are constants determined by the source orientation 
with respect to $\bm{e}_{+}$ and $\bm{e}_{\times}$.

For detectors that are small compared to the observed radiation 
wavelength\footnote{True for all terrestrial detectors. For LISA-like detectors, true for wave frequency $f\lesssim10$~mHz.} we may write the detector response to the 
incident $(h_{+}, h_{\times})$ as
\begin{align}
r(t) &= F_{+}(t)h_{+}(t,\vec{x}(t)) + F_{\times}(t)h_{\times}(t,\vec{x}(t))\,,
\label{e:response}
\end{align}
where $\vec{x}(t)$ is the detector location. 
The functions $F_{+}$ and $F_{\times}$ are determined by 
the projection of the detector's antenna pattern on the 
wave polarization tensors. Both terrestrial and space-based LISA-like detectors are constantly changing their orientation with respect to $\bm{e}_{+}$ and $\bm{e}_{\times}$ (in the case of terrestrial detectors owing to Earth's diurnal motion about its 
rotation axis, and in the case of space-based detectors 
owing to the science-craft constellation's annual motion about about Sol); correspondingly, 
$F_{+}$ and $F_{\times}$ are time dependent. 

\subsection{Known source location and frequency}

\begin{figure}[htbp!]
  \begin{center}
  \includegraphics[width=0.5\textwidth]{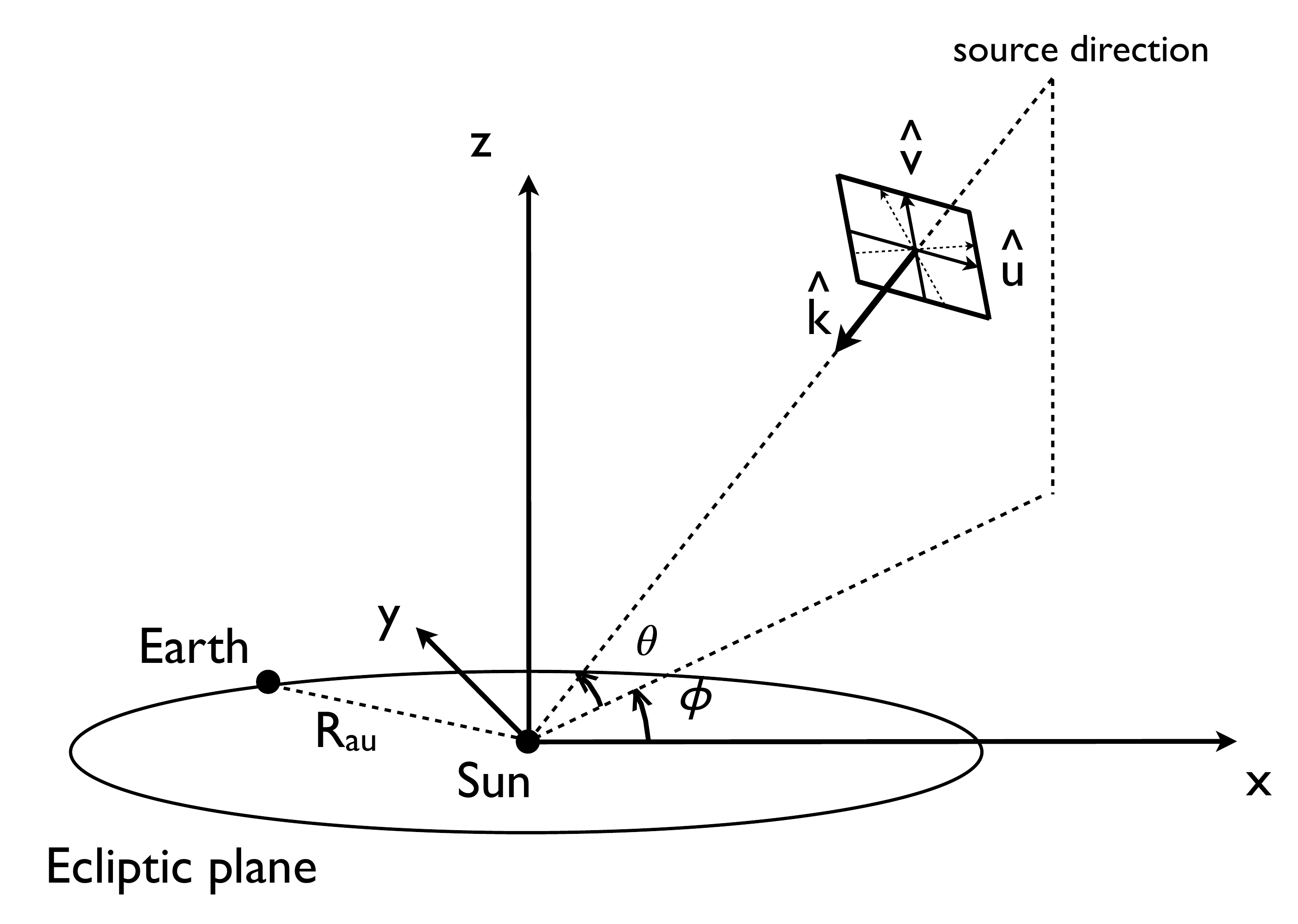}
  \caption{The relevant geometric quantities used 
  in the calculation:
  $\hat k$ is the unit vector pointing in the direction 
  of wave propagation;
  $\theta$ is the ecliptic latitude (i.e., the angle that 
  $-\hat k$ makes with the plane of the ecliptic);
  $\phi$ is the azimuthal angle of the source with respect
  to the Earth's orbital position at $t=0$.
  The detector antenna pattern functions 
  $F_+$ and $F_\times$ from Eq.~(\ref{e:response})
  are defined with respective to the polarization tensors 
  constructed from $\hat u$ and $\hat v$, which are 
  proportional to the unit vectors $\hat\phi$ and $\hat\theta$,
  respectively.}
  \label{f:geometry}
  \end{center}
\end{figure}

The Fisher Information matrix $\mathcal{I}$ has elements
\begin{align}
\mathcal{I}_{jk}(\vec\theta) &= \frac{2}{\sigma^2_n}\int_0^T 
\frac{\partial r}{\partial\theta_j}\frac{\partial r}{\partial\theta_k}dt\,,
\end{align}
where $T$ is the observation duration, 
$\sigma^2_n$ is the detector noise power spectral 
density at the gravitational wave frequency $f_{\mathrm{gw}}$, 
and $\vec{\theta}$ denotes the parameter vector 
$(\epsilon,H_{+}, H_{\times},\Phi_{+},\Phi_{\times})$. 
The partial derivatives of $r$ 
with respect our parameterization are
\begin{subequations}
\begin{align}
\frac{\partial r}{\partial\epsilon} &= 
-2\pi f_{\rm gw}\frac{\hat{k}\cdot\vec{x}}{\cem(1+\epsilon)^2}
\left\{
F_{+}H_{+}\sin\left[2\pi f_{\text{gw}}u+\Phi_{+}\right]+
F_{\times}H_{\times}\sin\left[2\pi f_{\text{gw}}u+\Phi_{\times}\right]
\right\}\,,
\label{e:drde}
\\
\frac{\partial r}{\partial H_{+}} &= 
F_{+}\cos\left(2\pi f_{\text{gw}}u+\Phi_{+}\right)\,,
\\
\frac{\partial r}{\partial H_{\times}} &= 
F_{\times}\cos\left(2\pi f_{\text{gw}}u+\Phi_{\times}\right)\,,
\\
\frac{\partial r}{\partial \Phi_{+}} &= 
-F_{+}H_{+}\sin\left(2\pi f_{\text{gw}}u+\Phi_{+}\right)\,,
\\
\frac{\partial r}{\partial \Phi_{\times}} &= 
-F_{\times}H_{\times}\sin\left(2\pi f_{\text{gw}}u+\Phi_{\times}\right)\,.
\end{align}
\end{subequations}
For all cases of interest the gravitational-wave detectors 
follow Earth's orbit about Sol; correspondingly, 
\begin{align}
\hat{k}\cdot\vec{x} &= \left(\rau\cos\theta\right)\cos(\omega_{\odot} t-\phi)\,,
\end{align}
where $\rau$ is Earth's orbital radius (1~au), 
$\omega_{\odot}$ is the detector angular velocity 
in its orbital motion about Sol ($2\pi/\text{yr}$), 
$\theta$ is the ecliptic latitude, 
and $\phi$ is the azimuthal angle of the source with respect
to Earth's orbital position at $t=0$.
(See Figure~\ref{f:geometry}.)
(The small displacement $\vec{d}$ of a terrestrial detector 
away from the Earth's orbital path about Sol introduces an order $d/\rau\sim0.25\%$ correction, which we ignore.)

To evaluate the components of the Fisher matrix we take advantage of the 
sinusoidal periodicity of $F_{+}$, $F_{\times}$, $h_{+}$ and $h_{\times}$ and focus on observations that are integer multiples of a year duration. Noting that 
\begin{align}
&\omega_{\odot}\ll2\pi f_{\text{gw}}\ll \cem/d\,,\\
&\omega_{\odot}\ll\cem/\rau\,,
\end{align}
the integrals for the Fisher matrix elements $\mathcal{I}_{\epsilon j}$ for $T>1\,\text{yr}$ quickly simplify to 
\begin{align}
\mathcal{I}_{\epsilon\epsilon} &= \frac{(2\pi f_{\text{gw}}\rau\rho\cos\theta)^2}{\cem^2(1+\epsilon)^4}\,,
\\
\mathcal{I}_{\epsilon j} &= 0\qquad\text{for $j\in\{H_{+},H_{\times},\Phi_{+},\Phi_{\times}\}$}\,,
\end{align}
where $\rho^2$ is the (power) signal-to-noise ratio
\begin{align}
\rho^2 &= \frac{1}{2\sigma_n^2}\int_0^T r^2(t)dt\,.
\end{align}
Correspondingly, at the level of the Cramer-Rao bound  
there is no co-variance between the uncertainty in our estimate of $\epsilon$ and 
any of the other problem parameters. The expected variance of the estimate for $\epsilon$ is thus
\begin{align}\label{eqn:result}
\nu_\epsilon &= \left(\mathcal{I}^{-1}\right)_{\epsilon\epsilon} \\
&= \frac{1}{\mathcal{I}_{\epsilon\epsilon}}=
\left(\frac{\cem}{2\pi f_{\rm gw}\rau}\right)^2\frac{\left(1+\epsilon\right)^4\sec^2\theta}{\rho^2}\,.
\end{align}
This result is valid for observations in all ground-based 
detectors and all proposed space-based detectors, 
whether Earth- or solar-orbiting. 
It is also valid for detector arrays where the data are combined 
coherently as described in, e.g., \cite{finn:2001:asf,finn:2010:dla}. 
(In the case of detector arrays $\rho^2$ is the \emph{array} 
power signal-to-noise ratio.) 
Details specific to the detectors, such as the antenna pattern 
functions $F_{+}$ and $F_{\times}$, cancel out when the uncertainty is expressed in terms 
of the signal-to-noise $\rho$.

\subsection{Unknown source location and/or frequency}
If the source frequency, sky location, or both are not known
\emph{a~priori} one needs to enlarge the parameter
vector $\vec\theta$ to include the additional unknowns: e.g., the frequency $f_{\text{gw}}$ and/or the source location on the sky ($\theta$, $\phi$). The Fisher matrix dimensionality thus expands to include terms involving partial derivatives
\begin{align}
\frac{\partial r}{\partial f_{\text{gw}}} &= 
-2\pi \left(
t - \frac{\hat k\cdot \vec x}{\cem(1+\epsilon)} 
\right)
\left\{
F_{+}H_{+}\sin\left[2\pi f_{\text{gw}}u+\Phi_{+}\right]+
F_{\times}H_{\times}\sin\left[2\pi f_{\text{gw}}u+\Phi_{\times}\right]
\right\}\,,
\\
\frac{\partial r}{\partial\theta} &= 
\frac{2\pi f_{\text{gw}}}{\cem(1+\epsilon)}
\frac{\partial(\hat k\cdot\vec x)}{\partial\theta}
\left\{
F_{+}H_{+}\sin\left[2\pi f_{\text{gw}}u+\Phi_{+}\right]+
F_{\times}H_{\times}\sin\left[2\pi f_{\text{gw}}u+\Phi_{\times}\right]
\right\}
\nonumber
\\
&\quad\quad
+ 
\left\{\frac{\partial F_{+}}{\partial\theta}\,H_{+}
\cos\left[2\pi f_{\text{gw}}u+\Phi_{+}\right]
+\frac{\partial F_{\times}}{\partial\theta}\,H_{\times}
\cos\left[2\pi f_{\text{gw}}u+\Phi_{\times}\right]
\right\}\,,
\\
\frac{\partial r}{\partial\phi} &= 
\frac{2\pi f_{\text{gw}}}{\cem(1+\epsilon)}
\frac{\partial(\hat k\cdot\vec x)}{\partial\phi}
\left\{
F_{+}H_{+}\sin\left[2\pi f_{\text{gw}}u+\Phi_{+}\right]+
F_{\times}H_{\times}\sin\left[2\pi f_{\text{gw}}u+\Phi_{\times}\right]
\right\}
\nonumber
\\
&\quad\quad
+ 
\left\{\frac{\partial F_{+}}{\partial\phi}\,H_{+}
\cos\left[2\pi f_{\text{gw}}u+\Phi_{+}\right]
+\frac{\partial F_{\times}}{\partial\phi}\,H_{\times}
\cos\left[2\pi f_{\text{gw}}u+\Phi_{\times}\right]
\right\}\,,
\end{align}
where
\begin{align}
\frac{\partial(\hat k\cdot\vec x)}{\partial\theta}
&=-\rau\sin\theta\,\cos(\omega_\odot t-\phi)\,,
\\
\frac{\partial(\hat k\cdot\vec x)}{\partial\phi}
&=+\rau\cos\theta\,\sin(\omega_\odot t-\phi)\,.
\end{align}
Comparing these expressions with the partial derivative
$\partial r/\partial\epsilon$ from Equation~\ref{e:drde},
one can see that the off-diagonal Fisher matrix elements 
$\mathcal{I}_{\epsilon j}$
for $j\in\{f_{\text{gw}},\theta,\phi\}$ are 
\emph{non-zero}. Correspondingly, the elements of the covariance matrix $(\mathcal{I}^{-1})_{\epsilon k}$ are no longer trivial and $\nu_\epsilon$ no longer simply expressed. How well we can estimate the propagation speed of gravitational waves using observations of periodic sources whose frequency or sky location are not known \emph{a priori} is the subject of work in-progress.



\section{Discussion}\label{sec:discussion}
As described here, to measure the gravitational-wave 
propagation speed from the R{\o}mer delay it is necessary to 
monitor a periodic source of gravitational waves, 
whose position on the sky and radiation frequency is known, 
for a year or more. 
For terrestrial detectors such a source might be a 
radio pulsar that also radiates gravitationally. 
For such sources, equation~\ref{eqn:result} may be 
written as
\begin{align}
\nu_\epsilon &= \left(3.2\times10^{-7}\right)^2
\left(\frac{100\,\text{Hz}}{f_{\text{gw}}}\right)^2
\left(\frac{10}{\rho}\right)^2\left(1+\epsilon\right)^4\sec^2\theta;
\end{align}
i.e., the ``$3\sigma$'' fractional uncertainty in the 
measurement of the gravitational-wave propagation speed 
arising from a one or more year observation of a 100~Hz 
gravitational-wave source situated on the ecliptic plane 
is $10^{-6}(10/\rho)$. 
Since a signal-to-noise $\rho\simeq10$ is typically taken 
as the threshold for source detection in a ground-based 
detector or detector network, if a periodic source is 
observed a measurement of $\cgw$ to $3\sigma$ precision,
$300(10/\rho)\,\text{m s}^{-1}$, will follow shortly. 

There are no reliable predictions for the gravitational-wave amplitude associated with rapidly rotating neutron stars. Mass asymmetries --- ``mountains'' --- are limited in size by the tensile strength of the neutron star crust, while the potential for fluid circulation instabilities (\emph{r}-modes) to lead to gravitational-wave emission depends on the (temperature dependent) neutron star surface ``ocean'' shear and bulk viscosities \cite[\S7.3]{sathyaprakash:2009:paa}. It may well be the case that neutron star crusts cannot support a sufficiently large asymmetry 
to be observable gravitational-wave sources, or that the neutron star fluid viscosity is always so great as to stabilize neutron star fluid \emph{r}-modes. Likewise, it may be that circumstances can be contrived that lead neutron stars to be strong radiation sources for ground-based detectors, but that there is no natural mechanism for creating or placing the neutron star into such states. 
Thus, while a sensitive measurement of the gravitational-wave speed is possible with ground-based 
detectors, carrying it out depends on the observation of a type of source that may not be available to us. 

Strong sources of periodic gravitational waves, 
in the form of galactic white dwarf binary systems, 
are both certain and plentiful for any of the proposed 
``LISA''-like space-based gravitational wave detectors \cite{danzmann:2011:luh}. 
For such sources, Equation~\ref{eqn:result} is conveniently
written as
\begin{align}
\nu_\epsilon &=
\left(3.2\times10^{-4}\right)^2
\left(\frac{10\,\text{mHz}}{f_{\text{gw}}}\right)^2
\left(\frac{100}{\rho}\right)^2\left(1+\epsilon\right)^4\sec^2\theta. 
\end{align}
An amplitude signal-to-noise of 100 in a one-year 
observation is the minimum expected for a typical 
``verification binary'' in a space-based detector; 
correspondingly, the ``$3\sigma$'' fractional uncertainty 
in the measurement of the gravitational-wave propagation 
speed arising from a one or more year observation of a 
10~mHz gravitational-wave source situated on the ecliptic 
plane is a quite respectable $10^{-3}(100/\rho)$.

At present, then, we find ourselves in an odd position. With the observation of a periodic gravitational-wave source, 
we know how to make a direct, accurate and unambiguous measure of the gravitational-wave propagation speed and, from it, a ``go/no-go'' test of general relativity theory.
On the one hand, for existing or foreseeable future ground-based detectors there is 
no guarantee that there will, or --- indeed --- can, exist any source that will enable the measurement. 
On the other hand, there are scores sources, already identified, that are accessible to a space-based detector that would enable 
such a measurement but, despite the strong recommendation of the United States National Research Council \cite{committee-for-a-decadal-survey-of-astronomy-and-astrophysics:2010:nwn}, NASA abandoned its committment to the decade-long ESA/NASA partnership that would have 
led to the construction of such an observatory and no such project is currently planned by either agency.
We can only hope that the feasibility of accurately and unambiguously testing 
general relativity --- by means such as described
here --- will strengthen the case for reviving a LISA-like gravitational wave observatory in the near future.


\begin{acknowledgments}
We gratefully acknowledge discussions with Martin Hendry and Graham Woan, and the hospitality of the Aspen Center for Physics. 
LSF acknowledges the support of NSF awards 0940924 and 0969857; 
JDR acknowledges the support of Leverhulme Trust RF/2005/0104 and
NSF award PHY-1205585.

\end{acknowledgments}



\end{document}